\documentclass[]{spie}  %>>> use for US letter paper

\usepackage{amsmath,amsfonts,amssymb}
\usepackage{graphicx}
\usepackage{siunitx, enumitem}
\usepackage[colorlinks=true, allcolors=blue]{hyperref}
 \usepackage{textcomp}
\usepackage{multirow}
\usepackage[section]{placeins}  % float barriers at section boundaries
\let\origsubsection\subsection
\renewcommand{\subsection}{\FloatBarrier\origsubsection}  % also barrier at subsections

\title{Fundamental Noise Limits of Infrared Detectors in the Presence of Readout Glow}
\author[a]{Michael Bottom}
\author[b]{Gustav Pettersson}
\author[c]{Pavaman Bilgi}
\author[c]{Guillaume Huber}
\author[a]{Charles-Antoine Claveau}
\author[b]{Nathan Lourie}
\author[b]{Robert Simcoe}
\author[d]{Shane Jacobson}

\affil[a]{Department of Astronomy, University of California, Berkeley; Berkeley, CA 94720, USA}
\affil[b]{MIT Kavli Institute for Astrophysics and Space Research, Cambridge, MA 02139, USA}
\affil[c]{Institute for Astronomy, University of Hawai`i, 2680 Woodlawn Dr, Honolulu, HI 96822, USA}
\affil[d]{NSF’s NOIRLab, Gemini North, 670 N. A’ohoku Place, Hilo, HI 96720, USA}
\authorinfo{Send correspondence to Michael Bottom, e-mail: mbottom@berkeley.edu}

\begin{document} 
\maketitle

\begin{abstract}
%1000 character version for SPIE police

%Read noise in infrared sensor arrays remains a major obstacle for ground- and space-based astronomy, and is a prohibitive constraint on both the extremely large telescopes and the Habitable Worlds Observatory. The main strategy for lowering it, averaging multiple non-destructive reads, gives far less improvement than the $1/\sqrt{N}$ scaling theory predicts. We show this can largely be explained by readout glow, photon emission from the sensor electronics that generates photoelectrons in the pixels during readout. Because glow accumulates with reads rather than averaging down, it imposes a fundamental noise floor of $\sigma_{\rm min} \approx 1.5\,\sigma_{RN}^{1/2}G^{1/4}$. This limits HxRG-like sensors to about 2-3 e- of noise, and linear-mode avalanche photodiodes to about 0.5 e-. We present laboratory data using both sensor architectures, with the LmAPD following the predicted limit to within 0.1 e- over two decades of averaging.

Read noise in infrared sensor arrays remains a major obstacle for ground- and space-based astronomy.  It has long been recognized that the upcoming extremely large telescopes cannot meet their full potential unless read noise is significantly improved, and it is also a prohibitive constraint on the Habitable Worlds Observatory, a space telescope with the goal of detection and characterization of nearby Earth-like exoplanets.

The main strategy for lowering read noise is averaging through multiple non-destructive reads.  However, this typically results in less noise reduction than the 1/$\sqrt{N}$ scaling predicted by theory.  In this work, we show the poor averaging behavior can largely be explained by readout glow, photon emission from the sensor electronics that generates photoelectrons in the pixels during readout.  Because glow accumulates with reads rather than averaging, this imposes a fundamental noise floor of $\sigma_{\rm min} \approx 1.5\,\sigma_{RN}^{1/2}G^{1/4}$.  This limits averaging in HxRG-like sensors to about 2-3e- of noise, and linear-mode avalanche photodiodes (LmAPDs) to about 0.5e-.  We present laboratory data using both sensor architectures, with the LmAPD following the predicted noise value to within 0.1 e- over two decades of averaging.
\end{abstract}


\section{INTRODUCTION}
\label{sec:intro}  % \label{} allows reference to this section
Infrared detectors are widely used in astronomy.  They are present in every major ground telescope, are key to enabling the spectacular success of the James Webb Space Telescope and are envisioned as playing a leading role in future space telescopes like the Habitable Worlds Observatory\cite{feinberg2026habitable}. Yet they substantially lag optical sensors in terms of noise.  New optical CMOS sensors and CCD ``skipper'' designs can resolve individual photon events with nearly no noise, and even the previous generation of CCDs had readout noise several times lower than infrared sensors.

It is recognized that the next generation of 30-meter class ground telescopes will not be able to reach their full potential if infrared detector noise does not improve\cite{sullivan_calibrated_2012}.  Equally seriously, the Habitable Worlds Observatory, NASA's next flagship observatory,  will not be able to fulfill its primary mission, characterizing extrasolar planets, if infrared detector noise does not reduce to near zero\cite{gaudi2020habitable, luvoir2019luvoir}.

Superconducting detectors are one avenue to achieving extremely low noise, and have fundamental advantages over semiconductor sensors, achieving effectively zero noise due to a band gap orders of magnitude smaller than the energy of an infrared photon.  However, they need to be cooled to near absolute zero and have yet to attain the multi-megapixel sizes of commercially available semiconductor sensors.

In semiconductor infrared sensors, the most serious noise source is due to the readout transistor in each pixel, known as ``read noise.''  This noise source is typically at least an order of magnitude higher than all others combined, and has not significantly improved in two decades.  A method of bypassing this is through avalanche photodiodes, where the signal photoelectrons are multiplied through a gain medium before readout.  However, the stochastic nature of the avalanche process introduces its own noise.

In both conventional and amplified sensors, the standard method of reducing read noise below that delivered by a single read is reading out the sensor non-destructively, without resetting it.  The output is then whatever electrons have accumulated in the pixel plus independent realizations of the Gaussian readout noise.  Thus, via the central limit theorem, the uncertainty on the mean number of electrons in the pixel reduces as the square root of the number of reads, or equivalently, the effective readout noise reduces by the same amount.

It may be expected that with a sufficient number of non-destructive reads, the effective read noise could be reduced indefinitely, or at least until the uncertainty on the pixel mean is well less than a single electron.  At that point, the true number of accumulated electrons could be determined unambiguously.\footnote{at 0.2 e- of effective read noise, the chance of rounding incorrectly (eg, measuring 10 e- when the true value is 11) is about 1\%.  At 1 e-, it is  more than 50\%}  This would be equivalent to a noiseless detector, e.g., the theoretical limit of performance. In practice, however, the effective noise flattens out after a few dozen reads, and even starts to rise, so that only a factor of about 2 to 4 improvement is attainable.  This has usually been attributed to low frequency drifts (i.e.\ 1/$f$ noise) in the detector that do not average down\cite{rauscherTeledyneH1RGH2RG2015, giardinoNIRSpecDetectorsNoise2012}.

In this work, we examine the contribution of readout glow to the inability to properly average down.  Readout glow is the phenomenon of photoelectrons generated by the detector readout electronics that accumulate in the pixel.  Glow was recently identified as the underlying reason for high dark current measurements in infrared sensors by Regan and Bergeron\cite{regan_zero_2020}, with Ives et al.\cite{ives2020characterisation}\ attributing and demonstrating a read noise floor caused by glow.  The origin was identified as photons generated inside the pixel source-follower and propagating from the ROIC to be absorbed in the HgCdTe\cite{pichon_pixel_2023}, though the exact physical mechanism for photon generation has not been conclusively determined. Here we extend this work to formally consider the effect of glow on commonly used readout averaging schemes.  We show that it can limit the inability to average down read noise effectively, and validate this with laboratory data from conventional and amplified sensors.

\section{Background}
\label{sec:bgd}
Infrared sensors have several noise sources\footnote{In this section, the conventional units of [e-/pixel/...] are used to avoid confusion, even though they should strictly be ``number of electrons''/pixel/...  so that correct unitless values may be combined like $RN^2 + DC\cdot t$}:

\begin{itemize}
\item \textbf{Dark current, or $DC$} is a Poissonian (per unit time) noise source that is typically described in units of [electrons/pixel/second or e-/px/s].  This is due to thermal charges generated in the semiconductor material itself, so more physically motivated units are current density [ampere/meter$^2$].  Typical dark current rates of scientific sensors are less than 0.001 e-/px/s.  

\item \textbf{Read noise, or $RN$} is a zero-mean Gaussian distribution that is generated every readout, due to the output MOSFET.  It is described in units of [electrons root-mean-squared or e- RMS], where the value is specified by the standard deviation of the distribution, and has a typical value of 10--30 e- RMS for infrared scientific sensors.

\item \textbf{Glow, or $G$} is a Poissonian (per read) noise source which is also generated each read, and is understood to be photons generated by the electronics and then absorbed in the substrate.  It is specified in units of [e-/px/read], and has typical values of 0.03--0.1 e-/px/read.\footnote{Here ``glow'' is what is more properly referred to as ``unit cell glow,'' which is intrinsic to each pixel readout, and is not related to things like ``multiplexer glow,'' which are essentially hot parts of the sensor that emit infrared radiation that is then absorbed.  That noise source is not fundamental to the sensor but has more to do with imperfect shielding.}

\item Reset, pedestal, or \textbf{KTC noise} is present every time the detector is reset, and is a zero-mean Gaussian with a standard deviation of $\sqrt{k_BT C}$, where the $T$ is the operating temperature and $C$ is the pixel capacitance.  It has units of [e- RMS], and at typical cryogenic temperatures and pixel capacitances it has a value of about 30-50 e- RMS.
\end{itemize}

The signal to noise ratio for one pixel when measuring a source of flux $f$ is

\begin{equation}
\mathrm{SNR}
=
\frac{f\,t_{\mathrm{exp}}}
{\sqrt{\,f\,t_{\mathrm{exp}}
\;+\; DC\,t_{\mathrm{exp}}
\;+\; G\,\dfrac{t_{\mathrm{exp}}}{t_{\mathrm{fr}}}
\;+\; \sigma_{RN}^{2}\,\dfrac{t_{\mathrm{exp}}}{t_{\mathrm{fr}}}
\;+\; \sigma_{KTC}^{2}\,}}
\label{eq:snr_conventional}
\end{equation}

\noindent where $t_{\mathrm{exp}}/t_{\mathrm{fr}}$ is the number of frames of length $t_{\mathrm{fr}}$ read out during the exposure.  The photon shot noise variance is included in the denominator as $f t_{\rm exp}$.

For a single frame, the standard deviations of the different noise terms above are $\sigma_{KTC}\sim$ 40, $\sigma_{RN}\sim15$, $\sqrt{DC \cdot t_{fr}} < 1$  (with $t_{fr}$ being frame time), and $\sqrt{G}<1$, with units of number of electrons.  It is clear that KTC noise dominates, followed by read noise, with the other terms being far smaller.  

Fortunately, it is possible to eliminate KTC noise by correlated double sampling (CDS), where the device is reset, read out, allowed to integrate the signal photons, read out again without resetting. By subtracting the first read from the second the read noise variance is doubled (so $\sigma_{RN}^{2}\rightarrow 2\sigma_{RN}^{2} = \sigma_{\rm CDS}^{2}$) but the KTC noise is perfectly subtracted. This ``CDS read noise'' or $\sigma_{CDS}$ is the usual value quoted for read noise in infrared detectors, but note that in this paper we frequently use $\sigma_{RN}$ which is the actual read noise and assumed to be related by $\sigma_{CDS} = \sqrt{2}\sigma_{RN}$.

This leaves read noise as the remaining large noise source.  To tackle read noise, various signal averaging schemes are used, the most common being Fowler sampling, where $N$ reads are taken at the beginning, $N$ at the end, and each of the two groups are averaged before being subtracted. CDS is thus equivalent to Fowler-1, and the read noise should effectively be reduced by $1/\sqrt{N}$ with Fowler-N this way. Another common readout mode is sample-up-the-ramp (SUTR) where N nondestructive reads are performed at a fixed rate and a line is fit to the slope of each pixel's values to measure the flux.

\section{Readout averaging with conventional sensors}
\label{sec:averaging}
Experience has shown that the read noise cannot be reduced indefinitely by signal averaging, and only factors of a few reduction are obtained.  The usual reasons given for this are $1/f$ type non-stationary noise, which prevents effective averaging.  In this section, we analyze the noise propagation of common readout averaging schemes, and examine how glow plays a role as well.  \textbf{The main point is that unlike read noise, glow does not average down since it is a Poissonian noise that increases with more reads.}  Dark current will also play a role, however, it is so low in modern sensors that we will ignore it in the following discussion.

In the following, the convention is for a random variable to be indicated by uppercase, and a realization of that random variable to be lowercase.  For example, the read noise $RN = \mathcal{N}(0, \sigma_{RN}^2)$, and $rn_1$ = 5, $rn_2$ = -3.

\begin{table}[h]
\centering
\caption{Signal and noise terms used in the readout model.}
\begin{tabular}{lccccccl}
\hline
\multirow{2}{*}{Type} & Random & \multirow{2}{*}{Distribution} & \multirow{2}{*}{Mean} & \multirow{2}{*}{Variance} & \multirow{2}{*}{Realization} & \multirow{2}{*}{Notes} \\
 & variable \\
\hline
Signal
& $F$
& $\mathrm{Poiss}(f\,t_{\mathrm{exp}})$
& $f\,t_{\mathrm{exp}}$
& $f\,t_{\mathrm{exp}}$
& $f$
& Photoelectron flux in e-/s \\

Dark Current
& $DC$
& $\mathrm{Poiss}(DC\,t_{\mathrm{exp}})$
& $DC\,t_{\mathrm{exp}}$
& $DC\,t_{\mathrm{exp}}$
& $dc$
& Thermally generated e-/s \\

Read Noise
& $RN$
& $\mathcal{N}(0,\sigma_{RN}^2)$
& $0$
& $\sigma_{RN}^2$
& $r_i$
& Independent per read \\

Glow
& $G$
& $\mathrm{Poiss}(G)$
& $G$
& $G$
& $g_i$
& Per-read glow electrons \\

KTC Noise
& $KTC$
& $\mathcal{N}(0,\sigma_{KTC}^2)$
& $0$
& $\sigma_{KTC}^2$
& $ktc_i$
& Uncertainty in reset level \\
\hline
\end{tabular}
\end{table}

For the following results in Fowler-N and SUTR, we validated the analytic forms against numerical simulations of detector pixels with various levels of read noise and glow, with excellent agreement. The results of the simulations may be found in the appendix.

\subsection{Correlated double sampling (Fowler-1)}
As mentioned in \ref{sec:bgd}, an effective way to remove KTC noise, the largest noise source, is through correlated double sampling (CDS), which eliminates KTC noise by measuring and subtracting the relevant realization of the noise through two non-destructive reads $s_1$ and $s_2$:
\begin{align}
s_1 &= ktc_1 + g_1 + r_1, \\
s_2 &= ktc_1 + f\cdot t_{\mathrm{exp}} + g_1 + g_2 + r_2
\end{align}
The CDS estimator is $\hat{S}_{CDS} \equiv s_2 - s_1$, yielding
\begin{align}
s_2 - s_1
    &= \left(ktc_1 + f\,t_{\mathrm{exp}} + g_1 + g_2 + r_2\right)
     - \left(ktc_1 + g_1 + r_1\right) \\
    &= f\,t_{\mathrm{exp}} + g_2 + (r_2 - r_1), \\
\mathbb{E}\!\left[\hat{S}_{CDS}\right]
    &= f\,t_{\mathrm{exp}} + G, \\
\mathrm{Var}\!\left[\hat{S}_{CDS}\right]
    &= f\,t_{exp} + G + 2\sigma_{RN}^2
\label{eq:cds_noise_and_mean}
\end{align}
The mean of this output is $F\cdot t_{\mathrm{exp}} + G$, and the variance is the summed variance of the photon shot noise, the glow, and the two read noise realizations.  The KTC noise is a fixed constant after the first reset, so it subtracts perfectly.  The same is true for the first glow photoelectron.  The variance of the read noise doubles, since two independent reads are subtracted, so the effective read noise is $\sqrt{2}$ higher. Since G is small compared to $\sigma_{RN}$ it can typically be neglected here.   %  Note dark current is ignored here because it is negligible compared to the other terms.

\subsection{Fowler-N sampling}
More than one non-destructive read may be taken at beginning and end, with the two groups then averaged and subtracted.  This is referred to as ``Fowler sampling,'' of which CDS is a special case (Fowler-1).  For example, Fowler-32 is 32 non-destructive reads taken right after the reset frame, and then 32 at the end.

The first group yields counts $s_i^{(1)}$ for each read $i$:
\begin{equation}
\label{eq:fowler_read_1}
s_i^{(1)} = ktc_1 + \sum_{k=1}^{i} \left[g_k^{(1)}\right] + r_i^{(1)},
\qquad i = 1,\dots,N .
\end{equation}
\noindent and the second yields
\begin{equation}
s_j^{(2)} = ktc_1 + f\,t_{\mathrm{exp}}
+ \sum_{k=1}^{N} \left[ g_k^{(1)}\right] 
+ \sum_{k=1}^{j} \left[ g_k^{(2)}\right] + r_j^{(2)},
\qquad j = 1,\dots,N .
\label{eq:fowler_read_2}
\end{equation}

Where the superscripts indicate the group; (1) for the first group and (2) for the second.  Notice that non-destructive read (i) includes glow contributions from every previous non-destructive read.  The mean and variance of the first group average are:
\begin{align}
\mathbb{E}\!\left[\bar{s}^{(1)}\right]
&= ktc_1 + \frac{N+1}{2}\,G,\\
\qquad 
\mathrm{Var}\!\left(\bar{s}^{(1)}\right)
&= G\,\frac{(N+1)(2N+1)}{6N}
+ \frac{\sigma_{RN}^2}{N}.
\end{align}
For the second group, they are
\begin{align}
\mathbb{E}\!\left[\bar{s}^{(2)}\right]
&= ktc_1 + f\,t_{\mathrm{exp}} + \frac{3N+1}{2}\,G  \\
\mathrm{Var}\!\left(\bar{s}^{(2)}\right)
&= f\,t_{\mathrm{exp}} + 
NG
+
G\,\frac{(N+1)(2N+1)}{6N}
+
\frac{\sigma_{RN}^2}{N}.
\end{align}
and the covariance between the two group averages is:
\begin{equation}
\mathrm{Cov}\!\left(\bar{s}^{(1)},\bar{s}^{(2)}\right)
=
\frac{N+1}{2}\,G.
\end{equation}
\noindent where we leave the details to the appendix.  We may then combine these to get the expected mean and variance of the Fowler-N estimator, $\hat{S} \equiv \bar{s}^{(2)} - \bar{s}^{(1)}$, where we have excluded the shot noise variance.

\begin{equation}
\boxed{
\begin{aligned}
\mathbb{E}\!\left[\hat{S}\right]
&= f\,t_{\mathrm{exp}} + NG \\
\mathrm{Var}\!\left[\hat{S}\right] 
&= 
%\frac{2G}{3}\left(N+\frac{1}{2N}\right)
%+
%\frac{2\sigma_{RN}^2}{N} = 
\left(\frac{2N}{3}+\frac{1}{3N}\right)G
+
\frac{2}{N}\sigma_{RN}^2
\end{aligned}
}
\label{eq:fowler_noise_and_mean}
\end{equation}

The estimator is biased, since the mean is not equal to the photon flux $f\,t_{\mathrm{exp}}$.  This is because the subtraction mostly removes the glow from the first group, but not the second.  The amount of bias can be rather substantial at low signal; for example a sensor with 0.3 e-/px/read would give 10 extra electrons of measured signal in Fowler-32 mode compared to CDS.  

A second consequence is that there is an optimal value $N=N_\mathrm{opt}$ that gives the minimum possible noise with Fowler sampling, $\sigma_\mathrm{min}$:

\begin{equation}
\boxed{
\begin{aligned}
N_{\mathrm{opt}}
&
\;\approx\;
1.7\, \frac{\sigma_{RN}}{\sqrt{G}}
\;\approx\;
1.2\, \frac{\sigma_{CDS}}{\sqrt{G}}
\\
\sigma_{\rm min} 
%= \sqrt{\mathrm{Var}_{\min}}
&\approx
1.5\,
\sigma_{RN}^{1/2}\,G^{1/4}
\;\approx\;
1.3\, \sigma_{CDS}^{1/2}\,G^{1/4}
\end{aligned}
}
\label{eq:fowler_optimum}
\end{equation}

This minimum noise depends weakly on glow (to the 1/4 power), so a factor of 10 improvement in glow only halves the minimum noise.  On the other hand, if glow truly is photons emitted from the pixel below the HgCdTe layer it should be possible to reduce it substantially with appropriate blocking layers.

\subsection{Sample-up-the-ramp}
In sample-up-the-ramp (SUTR), the data is fit with a line, and the flux rate $f$ is derived from the slope of the line fit.  This is distinct from Fowler, where the ultimate output is a count of the number of electrons, not a rate.  The mechanism of sample-up-the-ramp requires a two parameter fit to a line, but it is still meaningful to speak of the effective noise on the fit, so even if the outputs (a rate vs a count) do not even have the same units, one can estimate how read noise and glow will affect each in a consistent way by multiplying the count rate with the exposure time.

A full accounting of optimally sampling up the ramp is given in Brandt\cite{brandt_optimal_2024}, which includes analyzing arbitrary read patterns including the use of ``resultants,'' where intermediate groups of reads are averaged before fitting a line.  Brandt's paper also self-consistently estimates the covariance matrix from the data itself and corrects for the bias arising from this procedure.  For the purposes of this proceeding, we will restrict ourselves to the simple line-fit with evenly spaced, single read ``resultants'' and calculate the effects of glow on the derived parameters of SUTR data.

Let an exposure consist of $N$ non-destructive reads\footnote{Note here that $N$ refers to the total number of reads in the integration, whereas Fowler-$N$ refers to the number of reads in each of the two groups. I.e.\ a Fowler-1 (CDS) integration uses 2 total reads and corresponds to SUTR with N=2.} evenly spaced by the frame time $t_{\mathrm{fr}}$, with the $i$th read occurring at $t_i = i\,t_{\mathrm{fr}}$ after the initial reset at $t = 0$.  The total exposure time available for the line fit is $t_{\mathrm{exp}} = (N-1)\,t_{\mathrm{fr}}$; note that in SUTR, the exposure time is determined by the frame time and number of reads, unlike in Fowler where it is a free parameter.  The count rate is estimated by fitting a line $y = a\,t + b$ to the $N$ measurements and taking the slope $a$ as the rate.

It is clear that the effect of glow will basically cause the slope of the line (the input flux) to be overestimated due to the accumulation of electrons during the read. Explicitly, the counts will be
\begin{equation}
s_i = ktc_1 + f\,t_i + \sum_{k=1}^{i} g_k + r_i,
\qquad i = 1,\dots,N .
\label{eq:sutr_read}
\end{equation}

\noindent A least squares fit to this line ($y = a t + b$) will return a slope of $\mathbb{E}[\hat{a}] = f + G/t_{\mathrm{fr}}$, so the estimate of the total signal would be 

\begin{equation}
\mathbb{E}\!\left[\hat{S}_{\mathrm{SUTR}}\right] = f\,t_{\mathrm{exp}} + (N-1)\,G,
\label{eq:sutr_signal_mean}
\end{equation}

\noindent the SUTR analog of the Fowler-N count bias of $NG$.

The closed-form least-squares signal variance with read noise and photon shot noise, for $N$ evenly spaced reads, is given in the first equation by Rauscher\cite{rauscherDetectorsJamesWebb2007} (with $m=1$ assuming no intermediate group averaging):
\begin{equation}
\mathrm{Var}\!\left[\hat{S}_{\mathrm{SUTR}}\right] = \frac{12(N-1)}{N(N+1)}\,\sigma_{RN}^2 
+ \frac{6\,(N^2+1)}{5\,N(N+1)}\,f\,t_{\mathrm{exp}}
\label{eq:sutr_rauscher}
\end{equation}

\noindent For evenly spaced reads, glow has the same covariance structure as photon shot noise, as two reads share the accumulated Poisson events up to the earlier of them.  So one may simply substitute $f\,t_{\mathrm{exp}}$ with $f\,t_{\mathrm{exp}} + (N-1)\,G$ in Eq.~(\ref{eq:sutr_rauscher}). Again we drop the photon variance contribution and have:
\begin{equation}
\boxed{
\begin{aligned}
\mathbb{E}\!\left[\hat{S}_{\mathrm{SUTR}}\right]
&= f\,t_{\mathrm{exp}} + (N-1)\,G \\
\mathrm{Var}\!\left[\hat{S}_{\mathrm{SUTR}}\right] 
&= \frac{12(N-1)}{N(N+1)}\,\sigma_{RN}^2 
+ \frac{6\,(N^2+1)}{5\,N(N+1)}\,(N-1)\,G
\end{aligned}
}
\label{eq:sutr_noise_and_mean}
\end{equation}
Eq.~(\ref{eq:sutr_noise_and_mean}) reduces to the CDS variance ($2\sigma_{RN}^2 + G$) at $N=2$, consistent with Eq.~(\ref{eq:fowler_noise_and_mean}) at its corresponding Fowler-1 = CDS.

We may now compute the optimal number of samples for SUTR.  The equation is not easy to minimize analytically given the large numbers of ``N'' terms, but an approximation $(N-1)/(N+1) \approx 1$ and $(N^2+1)/N \approx N$ is valid for typical values of $N$ and allows the simplification
\begin{equation}
\mathrm{Var}\!\left[\hat{S}_{\mathrm{SUTR}}\right] \approx \frac{12\,\sigma_{RN}^2}{N} + \frac{6NG}{5}
\end{equation}
which may be more easily minimized to find:
\begin{equation}
\boxed{
\begin{aligned}
N_{\mathrm{opt}}^{\mathrm{SUTR}}
\;&\approx\; 3.2\,\frac{\sigma_{RN}}{\sqrt{G}}
\;\approx\; 2.2\,\frac{\sigma_{CDS}}{\sqrt{G}} \\
\sigma_{\min}^{\mathrm{SUTR}}
&\approx 2.8\,\sigma_{RN}^{1/2}\,G^{1/4}
\;\approx\; 2.3\,\sigma_{CDS}^{1/2}\,G^{1/4}
\end{aligned}
}
\label{eq:sutr_optimum}
\end{equation}

\section{Readout averaging with amplified sensors}
\subsection{Gain model and signal-to-noise}
The effects of glow become more important when considering amplified detectors such as linear-mode avalanche photodiodes (LmAPDs), a newer sensor technology with substantially lower readout noise.  In these sensors, the signal electrons are multiplied by a gain before the readout noise penalty.  The gain affects photoelectrons, glow electrons, and dark current differently.  The photoelectrons see the full gain $M$, the glow sees a gain of $cM$ where $c$ is about 0.5\footnote{we have tentatively measured this factor by examining jump heights at high gain}, and the dark current very little of it.  To clarify this, it is useful to look at the SNR equation (Eq. \ref{eq:snr_conventional}), first in output-referred and then input-referred mode:
%% Output-referred SNR with excess noise factors.
\begin{equation}
\mathrm{SNR}_{\mathrm{out}}
=
\frac{M\,f\,t_{\mathrm{exp}}}
{\sqrt{\,F\,M^2\,f\,t_{\mathrm{exp}}
\;+\; DC\,t_{\mathrm{exp}}
\;+\; F_g\,c^2 M^2\, G\,\dfrac{t_{\mathrm{exp}}}{t_{\mathrm{fr}}}
\;+\; \sigma_{RN}^2\,\dfrac{t_{\mathrm{exp}}}{t_{\mathrm{fr}}}\,}}
\label{eq:apd_snr_output}
\end{equation}

%% Input-referred SNR: divide numerator and denominator by M.
\begin{equation}
\mathrm{SNR}_{\mathrm{in}}
=
\frac{f\,t_{\mathrm{exp}}}
{\sqrt{\,F\,f\,t_{\mathrm{exp}}
\;+\; \dfrac{DC\,}{M^2}t_{\mathrm{exp}}
\;+\; F_g\,c^2\, G\,\dfrac{t_{\mathrm{exp}}}{t_{\mathrm{fr}}}
\;+\; \left(\dfrac{\sigma_{RN}}{M}\right)^2\,\dfrac{t_{\mathrm{exp}}}{t_{\mathrm{fr}}}\,}}
\label{eq:apd_snr_input}
\end{equation}

\noindent where the means and variances in the denominator have been modified following the usual rules of $\mathbb{E}[bX] = b\mathbb{E}[X]$ and $\mathrm{Var}[b X] = b^2 \mathrm{Var}[X]$ for a random variable $X$ and constant $b$.  The excess noise factor terms $F$ and $F_g$ (not to be confused with fluxes) are defined as the ratio of the second moment of the gain to its squared mean,
\begin{equation}
F = \frac{\mathbb{E}[A^2]}{\mathbb{E}[A]^2}
  = 1 + \frac{\mathrm{Var}[A]}{M^2},
\qquad
F_g = 1 + \frac{\mathrm{Var}[A_g]}{(cM)^2},
\label{eq:enf_definition}
\end{equation}
where $A$ and $A_g$ are the random gains seen by a signal photoelectron and a
glow electron respectively, with means $\mathbb{E}[A] = M$ and
$\mathbb{E}[A_g] = cM$, and are a simple way to account for the fact that the
gain is not a fixed constant, but has a distribution.  If gain were a
constant, $F$ would be equal to 1.

Using the second input-referred form of the SNR equation allows one to speak of ``effective'' noise terms, e.g., an effective read noise of $\sigma_{RN}/M$, an effective dark-current variance of $DC/M^2$, and a glow variance of $F_g\,c^2\,G$ per read.  Notice that if $M$ is increased substantially, and excess noise is near unity, the only remaining noise source in the denominator of Eq. \ref{eq:apd_snr_input} is glow. And indeed, it is this last glow term that sets the noise floor in amplified detectors, as we will show in the next section.

\subsection{Averaging with amplified detectors}
\label{sec:apd_averaging}
 
The results of Section \ref{sec:averaging} carry over to amplified detectors, for the most part.  The gain is a linear scaling, so the per-read glow realizations remain independent, the cumulative structure of the glow is unchanged, and the covariance between the Fowler groups survives.  The effective quantities defined above may simply be substituted into Eqs.~(\ref{eq:fowler_noise_and_mean}) and (\ref{eq:sutr_noise_and_mean}).
 
There is one subtlety.  The effective glow $G_{\mathrm{eff}} = F_g\,c^2\,G$ is a variance, and applies only to the variance expressions.  The glow bias however scales with the mean of the amplified glow, which is $c\,G$ per read: a constant multiple of a Poisson variable scales its mean by the constant but its variance by the constant squared.  With this in mind, the Fowler-N result for an amplified detector is
\begin{equation}
\boxed{
\begin{aligned}
\mathbb{E}\!\left[\hat{S}\right]
&= f\,t_{\mathrm{exp}} + N\,c\,G \\
\mathrm{Var}\!\left[\hat{S}\right]
&= \left(\frac{2N}{3}+\frac{1}{3N}\right) F_g\,c^2\,G
+ \frac{2\,\sigma_{RN}^2}{M^2 N}
\end{aligned}
}
\label{eq:apd_fowler_noise_and_mean}
\end{equation}
with the corresponding optimum following from Eq.~(\ref{eq:fowler_optimum})
under the same substitutions,
\begin{equation}
\boxed{
\begin{aligned}
N_{\mathrm{opt}}
&\;\approx\;
1.7\,\frac{\sigma_{RN}}{c\,M\sqrt{F_g\,G}}
\;\approx\
2.3\,\frac{\sigma_{CDS,\mathrm{eff}}}{\sqrt{G}}
\\
\sigma_{\rm min}
&\;\approx\;
1.5\,\sqrt{c}\;F_g^{1/4}\left(\frac{\sigma_{RN}}{M}\right)^{1/2} G^{1/4}
\;\approx\;
0.9\,\sigma_{CDS,\mathrm{eff}}^{1/2}\,G^{1/4}
\end{aligned}
}
\label{eq:apd_fowler_optimum}
\end{equation}

\noindent the approximations assume $c \approx 0.5$ and $F_g\approx1.1$, and note that $G$ is the output referred, not input referred glow.  Here $\sigma_{CDS, eff} = \sqrt{2}\sigma_{RN}/M$. The structure of these equations is preserved, but the optimal number of frames and minimum noise is reduced.  

The same substitutions in Eq.~(\ref{eq:sutr_noise_and_mean}) give the
amplified SUTR result,

%% ---------- SUTR optimum, with numerical approximation ----------
%% (c ~ 0.5, F_g ~ 1.1; sigma_RN,eff = sigma_RN / M)
\begin{equation}
\boxed{
\begin{aligned}
N_{\mathrm{opt}}^{\mathrm{SUTR}}
&\;\approx\; 3.2\,\frac{\sigma_{RN}}{c\,M\sqrt{F_g\,G}}
\;\approx\; 4.3\,\frac{\sigma_{CDS,\mathrm{eff}}}{\sqrt{G}}
\\
\sigma_{\min}^{\mathrm{SUTR}}
&\;\approx\; 2.8\,\sqrt{c}\;F_g^{1/4}\left(\frac{\sigma_{RN}}{M}\right)^{1/2} G^{1/4}
\;\approx\; 1.7\,\sigma_{CDS,\mathrm{eff}}^{1/2}\,G^{1/4}
\end{aligned}
}
\label{eq:apd_sutr_optimum}
\end{equation}
Setting $M, c, F_g = 1$ recovers the previous unamplified results.
 
There are a few interesting things to notice.  The key result is that the gain suppresses the read noise term as $1/M^2$, but the glow variance term $F_g\,c^2\,G$ contains no factor of $M$ at all (the fraction $c$ is independent of gain as far as we know).  \textbf{At high gain, glow is therefore the main noise floor of an amplified sensor}, and the minimum noise improves only as $M^{-1/2}$, with the weak $G^{1/4}$ dependence on the glow rate as before.  Also, in absolute terms $N_{\mathrm{opt}}$ shrinks as $1/M$: an amplified sensor reaches its glow-limited minimum after far fewer reads than a conventional one, so flattening and increasing of the curve happen faster.
 
Comparing the numerical forms of Eqs.~(\ref{eq:apd_fowler_optimum}) and (\ref{eq:apd_sutr_optimum}) with their unamplified counterparts, Eqs. ~(\ref{eq:fowler_optimum}) and (\ref{eq:sutr_optimum}), one can also see that the amplified detector reaches a noise floor about 25\% lower than the unamplified formulas would predict (0.9 prefactor vs 1.3).  This is because the glow sees only partial gain, its input-referred variance is suppressed by $F_g\,c^2 \approx 0.3$ relative to a fully amplified Poisson process.  The partial amplification of glow is thus a small unexpected benefit of the LmAPD architecture, on top of the read noise suppression that motivates it.  (Though of course, it would have been better to have no amplification of glow at all!)
 
The partial gain also produces a distinctive signature.  Input-referred, the
glow accumulates at a mean rate of $c\,G$ per read but contributes a variance
of $F_g\,c^2\,G$ per read, so its photon transfer curve slope (amusing because these are glow photons) is:
\begin{equation}
\frac{F_g\,c^2\,G}{c\,G} = c\,F_g ,
\label{eq:apd_fano}
\end{equation}
which is below unity for $c < 1/F_g$: the glow appears sub-Poissonian
when input-referred.  This can be used as a sanity check, since a fit to the signal increase per read measures $c\,G$, a fit to the noise-versus-$N$ curve (Eq. \ref{eq:apd_fowler_noise_and_mean}) measures $F_g\,c^2\,G$, and their ratio measures $c\,F_g$.

\subsection{Can current LmAPDs average down the single photon level?}
We can ask what gain is required for the effective noise to reach the deep sub-electron levels needed for photon number resolution, taken here to be $\sigma_{\rm pc} = 0.2$ effective electrons rms, giving a 99.5\% accuracy\cite{fossum2013modeling}.  Setting $\sigma_{\rm min} \le \sigma_{\rm pc}$ in Eq.~(\ref{eq:apd_fowler_optimum}) and solving for $M$,
\begin{equation}
M \;\gtrsim\; 2.3\,\frac{\sigma_{RN}\,\sqrt{F_g\,c^2\,G}}{\sigma_{\rm pc}^2}
\;\approx\; 56\;\sigma_{RN}\,\sqrt{F_g\,c^2\,G}
\qquad (\sigma_{\rm pc} = 0.2\ e^-),
\label{eq:apd_min_gain}
\end{equation}
with the SUTR requirement a factor of $(2.8/1.5)^2 \approx 3.5$ higher.  For example, a sensor with $\sigma_{RN} = 10$ e-, $c = 0.5$, $F_g = 1.1$, and $G = 0.01$ e-/pix/read requires only $M \gtrsim 30$ in Fowler mode, with approximately 22 reads.  Such gains are achievable with modest tunnel current.  However, recall that glow is a distribution of discrete events, amplified to a mean amplitude of $c \approx 0.5$ photon-equivalents; over many reads there is a substantial chance of a glow electron masquerading as a photon event, regardless of how far the Gaussian noise is averaged down.  To achieve single photon detection, it is far more profitable to simply read out up-the-ramp and look for jumps.  At high gain, if $\sigma_{RN}/(M\sqrt{N}) \lesssim 0.2$ e-, roughly, the jumps can be distinguished as changepoints from the unamplified dark current slope with high efficiency.  This readout mode is beyond the scope of this paper.

\section{Fundamental limits}
\begin{figure}[!h]
    \centering
    \includegraphics[width=0.8\textwidth]{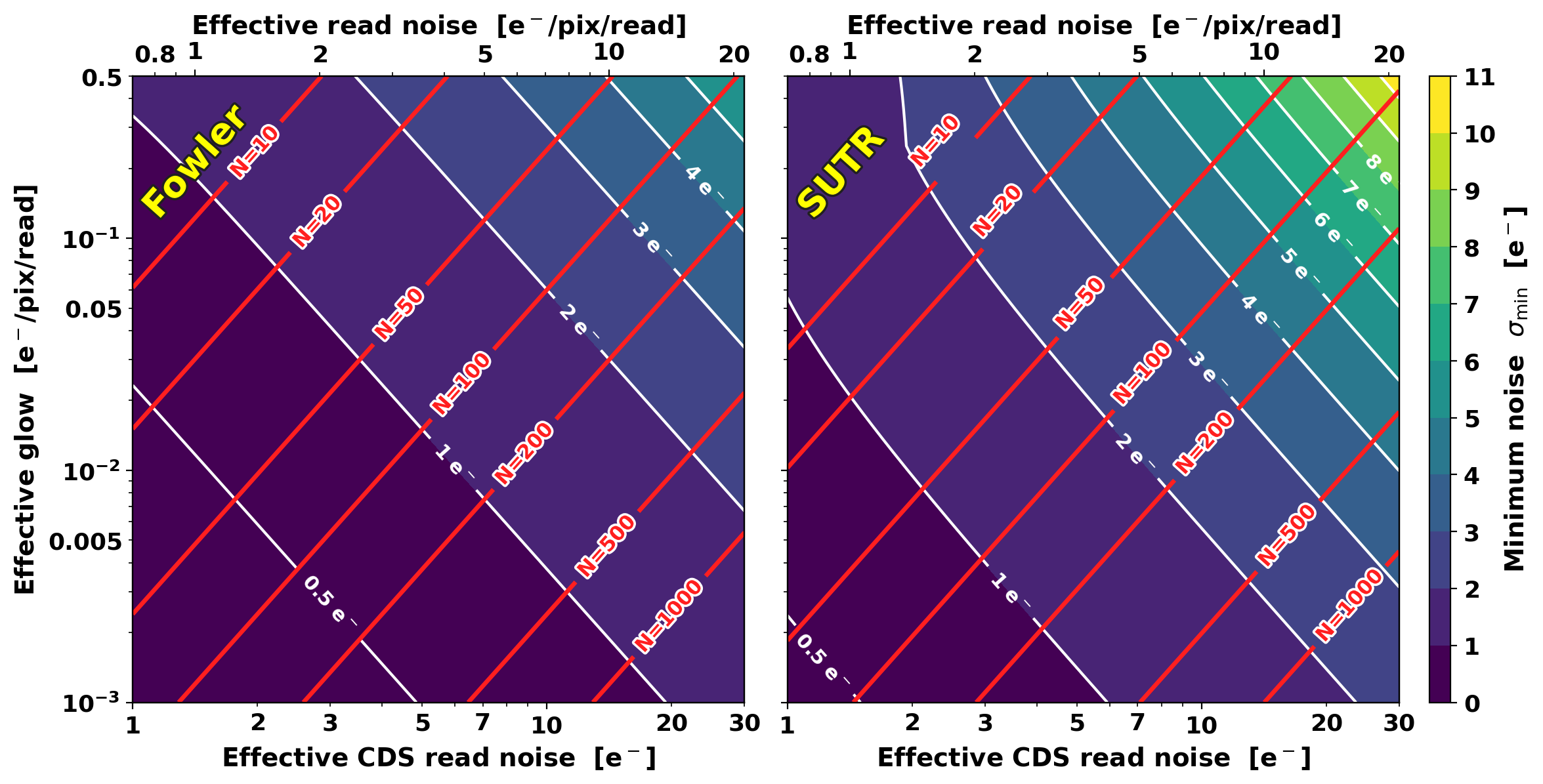}
    \caption{See text in this section for an explanation}
    \label{fig:minimum_noise_fowler_sutr}
\end{figure}

The plots show the fundamental noise limits from Eq.'s \ref{eq:fowler_optimum} and \ref{eq:sutr_optimum}.  The plots can be used to determine how much noise can be reasonably expected if all noise sources other than read noise and glow are eliminated.  For example, with a measured glow of 0.1 e-/pix/read, and a CDS read noise of 15 e-/pix/read, the minimum Fowler noise is about 3e- after N$\sim$100 reads (Fowler-50).  Improving glow to 0.01 e-/pix/read drops the minimum noise to $\sim$1.5 e-, but requires $\sim$300 reads to achieve.  Note that at low reads ($N<10$) where the approximations for high $N$ break down, the minimum is numerically determined, which accounts for the curvature seen on the log-log plot.  For amplified sensors, the plots are also valid, so long as the ``effective'' values of read noise and glow, $\sigma_{RN}/M$ and $F_g\,c^2\,G$, are used.

\section{Comparison with reality}

\subsection{FIRE H2RG}
To compare our theory with relevant data we performed tests on the science-grade Teledyne Hawaii-2RG (H2RG) detector in the Folded-port InfraRed Echellette (FIRE) spectrograph \cite{simcoe_FIRE_2013} at the \SI{6.5}{\m} Magellan Baade telescope at Las Campanas Observatory. Testing with an active science instrument is representative for real-world performance but has several limitations we have attempted to control.
\subsubsection{Methods}
The science detector in FIRE has 2048x2048 \SI{18}{\micro\m} pixels, a \SI{2.5}{\micro\m} cutoff HgCdTe photosensitive layer, and is operated at \SI{80}{\K} inside an instrument dewar that is held at approximately \SI{100}{\K}. The background in the instrument dewar is higher than the intrinsic dark current of the device and varies over time, especially during daytime when these data were collected. To minimize the background the ``blank'' slit was selected in the spectrograph but the dark counts were still upwards of \SI{.02}{e/px/s}, compared to the detector's intrinsic value that measured below \SI{.001}{e/px/s} during characterization in a dark dewar. To maximize the contribution of glow relative to this high background we used all 32 outputs from the device instead of the four that are used for science. This readout configuration yields a native frame time of $t_{fr}=\SI{1.4336}{\s}$ but the instrument's readout computer is not designed for this data rate, so we dropped frames to produce a half and a quarter of this rate. Since the background also varied over time we strategically timed the data collection so that slow variations could be removed by averaging across experiments. Another limitation is that the detector requires time to stabilize after the clocking pattern is changed, so we collected long ramps and only used a subset of frames at the end after the detector had stabilized.

The dataset was collected on 27 May 2026 from approximately 10am to noon local time and consisted of:
\begin{enumerate}[nosep]
    \item A 384-frame ramp at a rate of $t_{fr}=\SI{5.7}{\s}$ (3 drops), of which the last 128 frames were used. 
    \item A 375-frame ramp at a rate of $t_{fr}=\SI{2.9}{\s}$ (1 drop), of which the last 128 frames were used. 
    \item A 384-frame ramp at a rate of $t_{fr}=\SI{2.9}{\s}$ (1 drop), of which the last 128 frames were used. 
    \item A 384-frame ramp at a rate of $t_{fr}=\SI{5.7}{\s}$ (3 drops), of which the last 128 frames were used. 
\end{enumerate}
Following the method of Regan and Bergeron\cite{regan_zero_2020}, we can now separate the per-time background contribution from the per-read glow contribution in the FIRE detector from the change in total dark counts measured in these tests. The same data were used to measure the readout noise by combining the frames in the fast ramps to recreate Fowler sampling up to $N=64$ and SUTR up to $N=128$. With this method each ramp of 128 frames gave 64 independent Fowler-1 (CDS) measurements by subtracting the even frames from the odd, 32 independent Fowler-2 measurements by the same method after averaging adjacent pairs, etc. The final noise measurements and uncertainties were produced by taking the mean and standard deviation after rejecting outliers at the 5-sigma level. All results were scaled by the nominal gain of \SI{1.2}{e/DN}.

\begin{figure}[htb]
    \centering
    \includegraphics[width=1\linewidth]{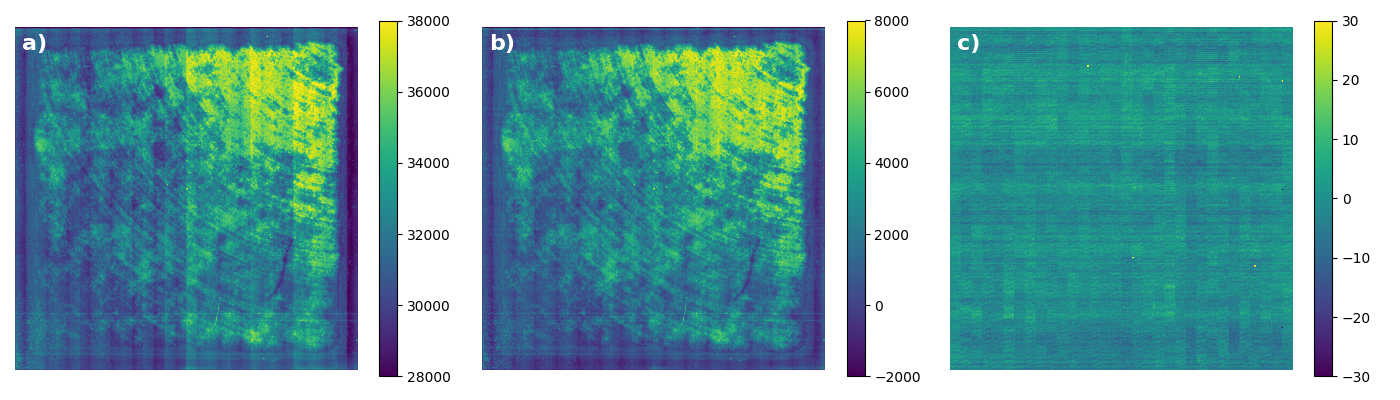}
    \caption{Stages of frame correction for FIRE H2RG data, all given in digital counts. a) Raw frame. b) Reference pixel corrected frame. c) Correlated double sample (CDS) of corrected frames.}
    \label{fig:h2rg_corrections}
\end{figure}

Before any analysis all raw frames were corrected using the reference pixels (the outer four rows and columns) on the detector. First, the per-amplifier offset and alternating column noise were removed by averaging and subtracting all reference pixels belonging to each output, treating odd and even columns separately. Second, fast drifts were corrected by averaging and subtracting the row reference pixels in blocks of 16 rows. In both cases outliers were rejected at the 5-sigma level before taking the mean. Figure~\ref{fig:h2rg_corrections} shows a comparison between a raw frame, a reference pixel corrected frame, and a CDS frame that also removes the fixed ``tree ring'' bias pattern and KTC reset noise. The CDS frame shows that each output has a slightly different noise characteristic (vertical bands) and that most but not all temporal noise (horizontal stripes) is removed by the simple row-block reference pixel correction.

\subsubsection{Results}
\begin{figure}[htb]
    \centering
    \includegraphics[width=.8\linewidth]{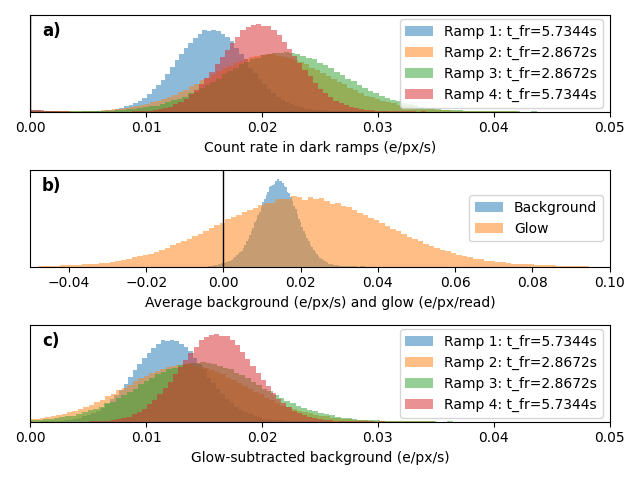}
    \caption{4x4 pixel average histograms of the FIRE H2RG data. a) Raw count rates. b) The average contributions of readout glow and the instrument's background separated. c) Background with a glow of \SI{.021}{e/px/read} subtracted.}
    \label{fig:h2rg_dark_glow_histograms}
\end{figure}
Figure~\ref{fig:h2rg_dark_glow_histograms}a shows histograms of the count rates in the ramps collected for this work. As the detector is read out more frequently the count rate increases, which confirms the presence of readout glow. It is also clear that the background in the instrument increased over time, which makes these results susceptible to systematic errors.
\begin{figure}[htb]
    \centering
    \includegraphics[width=1\linewidth]{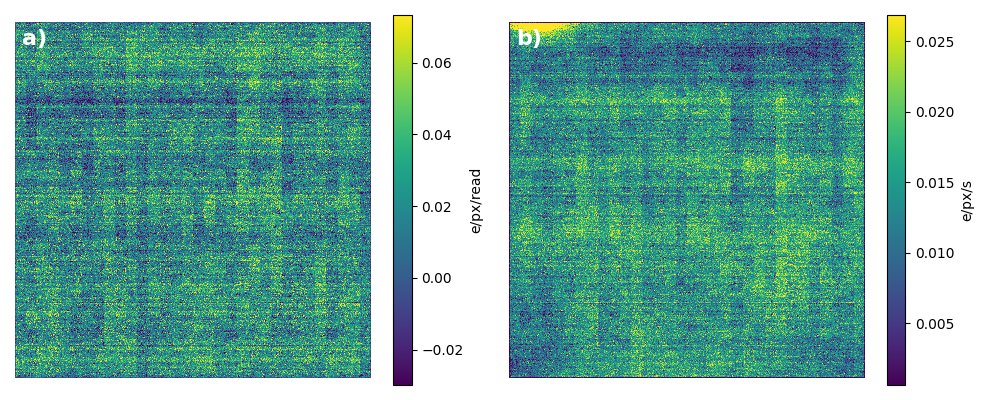}
    \caption{4x4 pixel average per-read glow (a) and per-time background counts (b) for the FIRE H2RG.}
    \label{fig:h2rg_glow_dark_maps}
\end{figure}
In Figure~\ref{fig:h2rg_dark_glow_histograms}b the glow and background for the FIRE H2RG data have been separated after averaging the fast and slow ramps; the spatial distribution of the data is given in Figure~\ref{fig:h2rg_glow_dark_maps}. Unlike Regan and Bergeron's results,\cite{regan_zero_2020} there is no discernible pattern in the glow of this detector. The background in the dewar is also even except for a bright spot in one corner which is typical for FIRE data that we believe originates somewhere in the ROIC outside the pixel area. With the contribution from glow subtracted in Figure~\ref{fig:h2rg_dark_glow_histograms}c we measure that the background rate increased from approximately \SIrange{.012}{.016}{e/px/s} during the test.

For the FIRE H2RG detector we measure the glow to \SI{.021}{e/px/read} and estimate the uncertainty to be about \SI{.01}{e/px/read} due to the varying background. This detector exhibits less glow than the \SI{.076}{e/px/read}\cite{regan_zero_2020} for Regan and Bergeron's \SI{5}{\micro\m} cutoff H2RG and is similar to the \SI{.03}{e/px/read}\cite{ives2020characterisation} for Ives et al.'s \SI{2.5}{\micro\m} cutoff H4RG-15. As these three devices were manufactured at different times and are operated in different ways it's unclear how comparable the results are, but if glow is caused by photons emitted from the ROIC absorbed in the HgCdTe as currently understood\cite{pichon_pixel_2023}, one would expect that devices sensitive to a broader range of photons exhibit more glow as we see here.

\begin{figure}[htb]
    \centering
    \includegraphics[width=.8\linewidth]{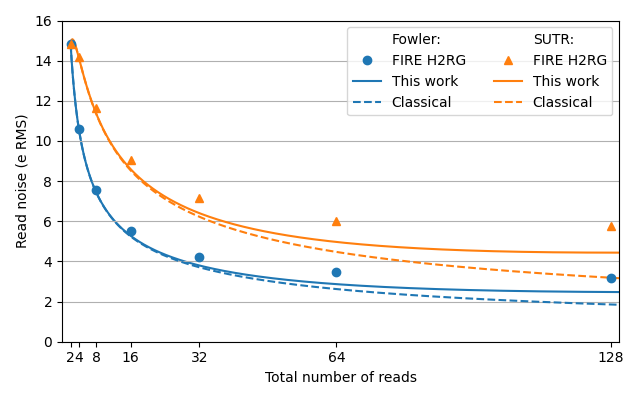}
    \caption{Measured and predicted noise for the first output of the FIRE H2RG detector. Note that there are $2N$ total reads for Fowler-N. The statistical uncertainty is smaller than the markers for all measurements, approximately 1\%.}
    \label{fig:h2rg_noise_results}
\end{figure}
The measured read noise for the FIRE H2RG is compared to theory in Figure~\ref{fig:h2rg_noise_results}. The lines for each averaging strategy compare the ``classical'' scaling that assumes independent measurements without glow or dark current, scaling as $\sqrt{\frac{1}{N}}$ for Fowler-N and $\sqrt{6\frac{N-1}{N(N+1)}}$ for SUTR-N, with the theory presented in this work. We use Eqs.~\ref{eq:fowler_noise_and_mean} and \ref{eq:sutr_noise_and_mean} with $G=\SI{.021}{e/px/read}$ and in both cases also include shot noise from a dark current of $DC=\SI{.014}{e/px/s}$ in the predictions. In this case the noise contribution from dark current is higher than that for glow, confirming that glow is not limiting in this particular instrument. The measured noise is higher than our theory predicts and at least some of that difference could be explained by the structured noise we see in the CDS frame in Figure~\ref{fig:h2rg_corrections}c. The CDS read noise per amplifier varied from 15 to 20 e RMS but the scaling with multiple samples was similar in all outputs.

\subsection{Ike Pono LmAPD}
\label{sec:ikepono}

We now compare the predicted averaging behavior against laboratory measurements of a linear-mode HgCdTe avalanche photodiode (LmAPD). Unlike the conventional FIRE H2RG detector above, this device multiplies the signal before the read-noise penalty, providing a test of the model on an amplified sensor.

\subsubsection{Methods}

We performed tests on a science-grade Leonardo ``Ike Pono'' linear-mode HgCdTe electron-avalanche photodiode array built on the ME1071 readout integrated circuit, read out through a SIDECAR ASIC and MACIE controller on the test bench developed by Claveau et al.\cite{claveau_first_2022}. The background level in the dewar is less than 4 photoelectrons per pixel per day. The 16 outputs of the device were read out at a pixel rate of 100\,kHz, corresponding to a frame time of $t_{fr}=\SI{0.69}{\s}$, with the detector regulated to 70\,K to within 1\,mK. We analyze an $80\times256$-pixel region ($20{,}480$ pixels) and scaled to input photoelectrons using the measured gain at each bias. We acquired data at biases from 6 to 14\,V, corresponding to gains of $\sim$2.5-38. Since tunnelling current (a dramatic increase in the dark current rate) sets in above 10\,V, the analysis in this section is restricted to the tunnelling-free regime at 10\,V and below. For simplicity we focus on 10\,V in what follows, although we obtained results at every bias voltage.

We tested the Fowler and SUTR averaging schemes on a single dataset. At each bias we recorded 100 ramps of 128 reads, each begun from a reset, and formed both schemes from these same ramps: SUTR-$N$ by least-squares fitting a line to the first $N$ reads (up to $N=128$), and Fowler-$N$ by differencing the average of the first $N$ reads from the average of the next $N$ reads (up to $N=64$, i.e.\ 128 total reads). We take the effective noise at each $N$ as the standard deviation across the 100 ramps. We also measured the dark current and glow directly, following the method of Regan and Bergeron\cite{regan_zero_2020}.  We recorded a fast ramp reading out continuously over roughly two hours and a slow ramp reading out once per 100 frame times over roughly four hours (a factor of 100 in read rate). From the fast ramp we additionally extracted a direct, fit-free read noise as the correlated double sample of adjacent reads: for each pair of consecutive reads we differenced the two frames over the region of interest and estimated $\sigma_{RN}$ as the robust standard deviation of that difference divided by $\sqrt{2}$. All frames were reference-corrected using the detector's reference rows, and statistics were computed after 5-sigma outlier rejection.

Because the data are scaled to input photoelectrons, the quantities returned by the fit are the input-referred ones of Section~\ref{sec:apd_averaging}.  The amplified variance expressions follow from Eqs.~(\ref{eq:fowler_noise_and_mean}) and~(\ref{eq:sutr_noise_and_mean}) under the substitutions $\sigma_{RN}\to\sigma_{RN}/M$ and $G\to F_g\,c^2\,G$ (Eq.~\ref{eq:apd_fowler_noise_and_mean}); the functional form is unchanged, so the fit is identical and simply returns the effective read noise $\sigma_{RN}/M$ and the glow variance $F_g\,c^2\,G$ in place of $\sigma_{RN}$ and $G$.  We sampled the two-parameter posterior with a Markov-chain Monte Carlo sampler\cite{foreman-mackey_emcee_2013}, weighting each point by its sampling uncertainty from the 100 ramps, and quote posterior medians and $1\sigma$ credible intervals.
\subsubsection{Results}

We begin with the ``direct'' glow measurement, at 10\,V gain, using the method of Regan and Bergeron\cite{regan_zero_2020}. This measurement is a fit to the signal accumulated per read, so for an amplified detector we solve for the input-referred mean glow $cG$, i.e. the mean glow term in Eq.~(\ref{eq:apd_fowler_noise_and_mean}). We find that $cG = 7.6\times10^{-3}$~e$^-$/pix/read and, from the same fit, a dark current of $\mathrm{DC} = 1.2\times10^{-4}$~e$^-$/pix/s, consistent with the (glow-subtracted) dark current of $\approx1\times 10^{-4}$~e$^-$/pix/s measured by Claveau et al.\cite{claveau_first_2022} at a slightly lower temperature.
\begin{figure}[htb]
    \centering
    \includegraphics[width=0.8\linewidth]{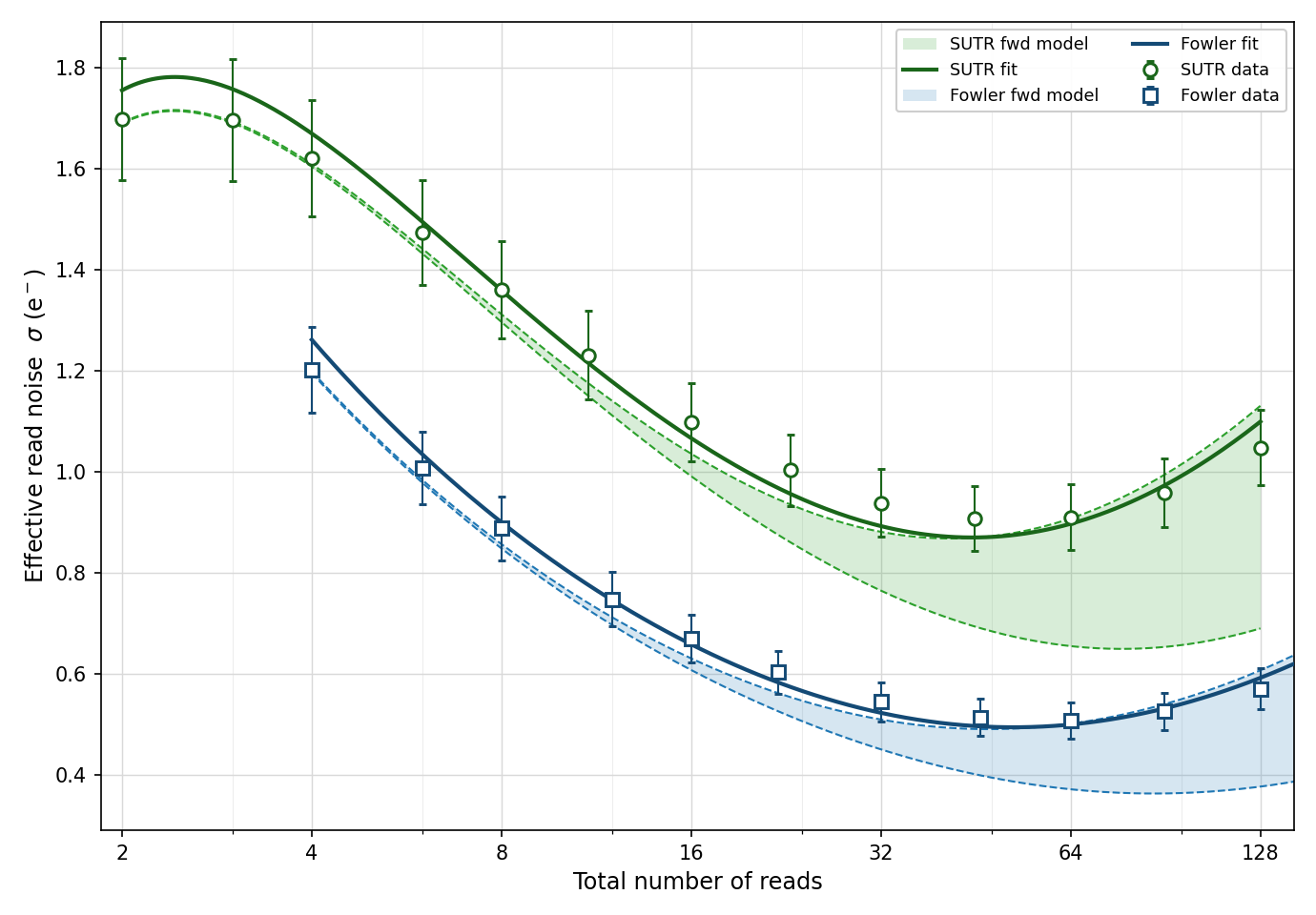}
    \vspace{6pt}
    \caption{Effective read noise versus the total number of reads at 10\,V bias for the Ike Pono LmAPD (Fowler-$N$ at $2N$ total reads, SUTR-$N$ at $N$). Points are the measured Fowler and SUTR noise. Solid curves are the two-parameter fits of the amplified noise model of Section~\ref{sec:apd_averaging}. The shaded wedge is the forward model built from the directly measured mean glow $cG$, with the glow variance $F_g\,c^2\,G$ swept over $c\,F_g = 0.3$ to $1.0$ (partially amplified to Poissonian). The read noise is held at its directly measured value.}
    \label{fig:ikepono_noise}
    \vspace{4pt}
\end{figure}

We next test the averaging behavior itself. Figure~\ref{fig:ikepono_noise} shows the effective read noise as a function of $N$ at 10\,V for both schemes. In each case the measured noise falls with $N$, reaches a minimum, and rises again at large $N$ once the glow term dominates, as predicted by Eqs.~(\ref{eq:apd_fowler_optimum}) and~(\ref{eq:apd_sutr_optimum}). The solid curves are fits of the amplified noise model of Section~\ref{sec:apd_averaging}, with the read noise and glow as the two free parameters. The glow returned by these fits is the effective glow variance $F_g\,c^2\,G$, the glow term of the noise variance (Eq.~\ref{eq:apd_fowler_noise_and_mean}). This differs from the mean glow $cG$ returned by the direct measurement. For an unamplified Poisson glow the mean and variance would coincide, but the partial amplification separates them, the mean scaling with $c$ and the variance with $c^2 F_g$, so that their ratio is $c\,F_g$ (Eq.~\ref{eq:apd_fano}). The fitted parameters are listed in Table~\ref{tab:ikepono_glow}.The measured mean glow $cG$ can also be used in the forward direction, propagated through the same amplified models with glow variance $F_g\,c^2\,G = (c\,F_g)\,(cG)$, to predict a noise curve for comparison with our fits.

\begin{table}[htb]
\centering
\vspace{4pt}
\caption{Read noise, glow, and dark current for the Ike Pono LmAPD at 6, 8, and 10\,V from the two averaging-scheme fits and the direct measurement. Both averaging schemes are formed from the same dataset. The noise fits solve for the glow variance $F_g\,c^2\,G$ (Eq.~\ref{eq:apd_fowler_noise_and_mean}), whereas the direct measurement gives the input-referred mean glow $cG$. The $G$ row gives the underlying glow backed out from each column, assuming $c = 0.5 \pm 0.1$ and $F_g = 1.1 \pm 0.1$ (Section~\ref{sec:apd_averaging}).}
\label{tab:ikepono_glow}
\vspace{6pt}
\renewcommand{\arraystretch}{1.35}
\setlength{\tabcolsep}{9pt}
\begin{tabular}{lccc}
\hline
 & SUTR & Fowler & Direct \\
\hline
\multicolumn{4}{l}{\textbf{6 V}} \\
$\sigma_{\mathrm{RN, eff}}$ (e$^-$)                 & $4.33\pm0.10$ & $4.37\pm0.12$ & $4.14\pm0.05$ \\
$F_g\,c^2\,G$ ($10^{-3}$~e$^-$/pix/read)  & $20.61\pm3.55$ & $21.41\pm4.03$ & $-$ \\
$cG$ ($10^{-3}$~e$^-$/pix/read)           & $-$ & $-$ & $9.82\pm0.05$ \\
$G$ ($10^{-3}$~e$^-$/pix/read)            & $74.9\pm33.3$ & $77.9\pm35.1$ & $19.6\pm3.9$ \\
DC ($10^{-4}$~e$^-$/pix/s)                      & $-$ & $-$ & $1.99\pm0.37$ \\
\hline
\multicolumn{4}{l}{\textbf{8 V}} \\
$\sigma_{\mathrm{RN, eff}}$ (e$^-$)                 & $2.35\pm0.06$ & $2.37\pm0.07$ & $2.25\pm0.03$ \\
$F_g\,c^2\,G$ ($10^{-3}$~e$^-$/pix/read)  & $10.41\pm1.40$ & $10.56\pm1.58$ & $-$ \\
$cG$ ($10^{-3}$~e$^-$/pix/read)           & $-$ & $-$ & $8.42\pm0.04$ \\
$G$ ($10^{-3}$~e$^-$/pix/read)            & $37.9\pm16.3$ & $38.4\pm16.8$ & $16.8\pm3.4$ \\
DC ($10^{-4}$~e$^-$/pix/s)                      & $-$ & $-$ & $1.42\pm0.34$ \\
\hline
\multicolumn{4}{l}{\textbf{10 V}} \\
$\sigma_{\mathrm{RN, eff}}$ (e$^-$)                 & $1.24\pm0.03$ & $1.26\pm0.04$ & $1.19\pm0.01$ \\
$F_g\,c^2\,G$ ($10^{-3}$~e$^-$/pix/read)  & $7.06\pm0.71$ & $7.05\pm0.76$ & $-$ \\
$cG$ ($10^{-3}$~e$^-$/pix/read)           & $-$ & $-$ & $7.58\pm0.04$ \\
$G$ ($10^{-3}$~e$^-$/pix/read)            & $25.7\pm10.8$ & $25.6\pm10.9$ & $15.2\pm3.0$ \\
DC ($10^{-4}$~e$^-$/pix/s)                      & $-$ & $-$ & $1.23\pm0.19$ \\
\hline
\end{tabular}
\end{table}

The two glow measurements, the mean glow $cG$ from the direct measurement and the glow variance $F_g\,c^2\,G$ from the fit to the noise-versus-$N$ curve, are related, with Eq.~(\ref{eq:apd_fano}) giving their ratio as $(F_g\,c^2\,G)/(cG) = c\,F_g$. This ratio is below unity when the glow is only partially amplified (Section~\ref{sec:apd_averaging}). Having tentatively measured the partial-gain fraction to be $c \approx 0.5$, we would expect the glow's excess noise factor $F_g$ to be in the range of 1.1-2. However, as we do not measure $F_g$ directly, we instead vary $c\,F_g$ over the range $0.3$ to $1.0$, spanning partially amplified to fully Poissonian. Propagating the directly measured mean glow through the amplified noise model over this range gives the shaded wedge in Fig.~\ref{fig:ikepono_noise}. The measured noise lies close to this wedge, tracking it most closely at low to moderate $N$, indicating that the independently measured glow captures much of the observed averaging behavior. With $c = 0.5$ and $F_g = 1.1$ (Section~\ref{sec:apd_averaging}), each measurement can also be converted to the underlying glow $G$, shown in the $G$ row of Table~\ref{tab:ikepono_glow}. The two averaging schemes agree; the value inferred from the direct glow is lower.

\section{Conclusion}
In this work, we have identified glow as a fundamental limit to how far noise can be averaged down in conventional and amplified infrared detectors.  Because glow accumulates with every read rather than averaging down, each averaging scheme has an optimal read count and a noise floor, $N_{\rm opt} \approx 1.7\,\sigma_{RN}/\sqrt{G}$ and $\sigma_{\rm min} \approx 1.5\,\sigma_{RN}^{1/2}G^{1/4}$ for Fowler, with
coefficients of 3.2 and 2.8 for SUTR.  The floor is only a 1/4-power function of glow, so efforts to reduce noise by reducing glow will succeed only if they reduce it substantially.  

We compared this behavior to real-world data, both by independently measuring glow and read noise and by deriving them from the data themselves.  While glow is not the limiting noise in our test of an H2RG in FIRE, owing to the background, it still contributes to the fundamental noise floor for this family of detectors.  In our LmAPD tests the measured noise follows the limit set by glow to within 0.1 e- over two decades of averaging, bottoming out near 0.5 e- at 10\,V bias against a read noise of 1.2 e-.  For that device the glow, rather than 1/$f$-type drift, accounts for the observed averaging behavior and for the inability to reach truly deep sub-electron noise by averaging alone.

The glow inferred from the noise fits sits a bit above the directly measured value at every bias voltage, consistent within the uncertainties but persistently offset.  Resolving this requires a direct measurement of the glow excess noise factor $F_g$ and the fractional gain factor $c$, which currently dominate the uncertainty in converting either measurement into a physical glow rate.  These are challenging to measure accurately as the absolute level of glow is quite low.

As a diagnostic, it would be prudent for practitioners to independently characterize glow and read noise, then derive the fundamental limits for their chosen averaging scheme; comparing the two informs how much noise is left on the table.  Finally, we expect this behavior in any sensor that is read
non-destructively, and encourage attempts to measure and characterize glow in such devices.
\acknowledgements
We acknowledge the engineering team at Leonardo UK for helpful discussions, and Luc Boucher and his group at ESA for many productive exchanges on glow. MB, PB, GH, CAC, and SJ were supported by NASA grant 18-SAT18-0028.  GP, NL, and RS were supported by The Kavli Foundation via grant PS-2025-GR-0237-3087.

\appendix    %>>>> this command starts appendixes

\section{Numerical validation}
We performed simulations to generate artificial detector measurements to validate our analytical results. In the simulation below, 10,000 realizations of Fowler sampling and SUTR were produced per data point for a detector with no dark current, a read noise of 11.1 e- RMS (16 in CDS), and with glows ranging from 0.3 to 0.03 e-/px/read. These show excellent agreement with Eqs.\ \ref{eq:fowler_noise_and_mean} and \ref{eq:sutr_noise_and_mean}.

\begin{figure}[!h]
    \centering
    \includegraphics[width=0.8\textwidth]{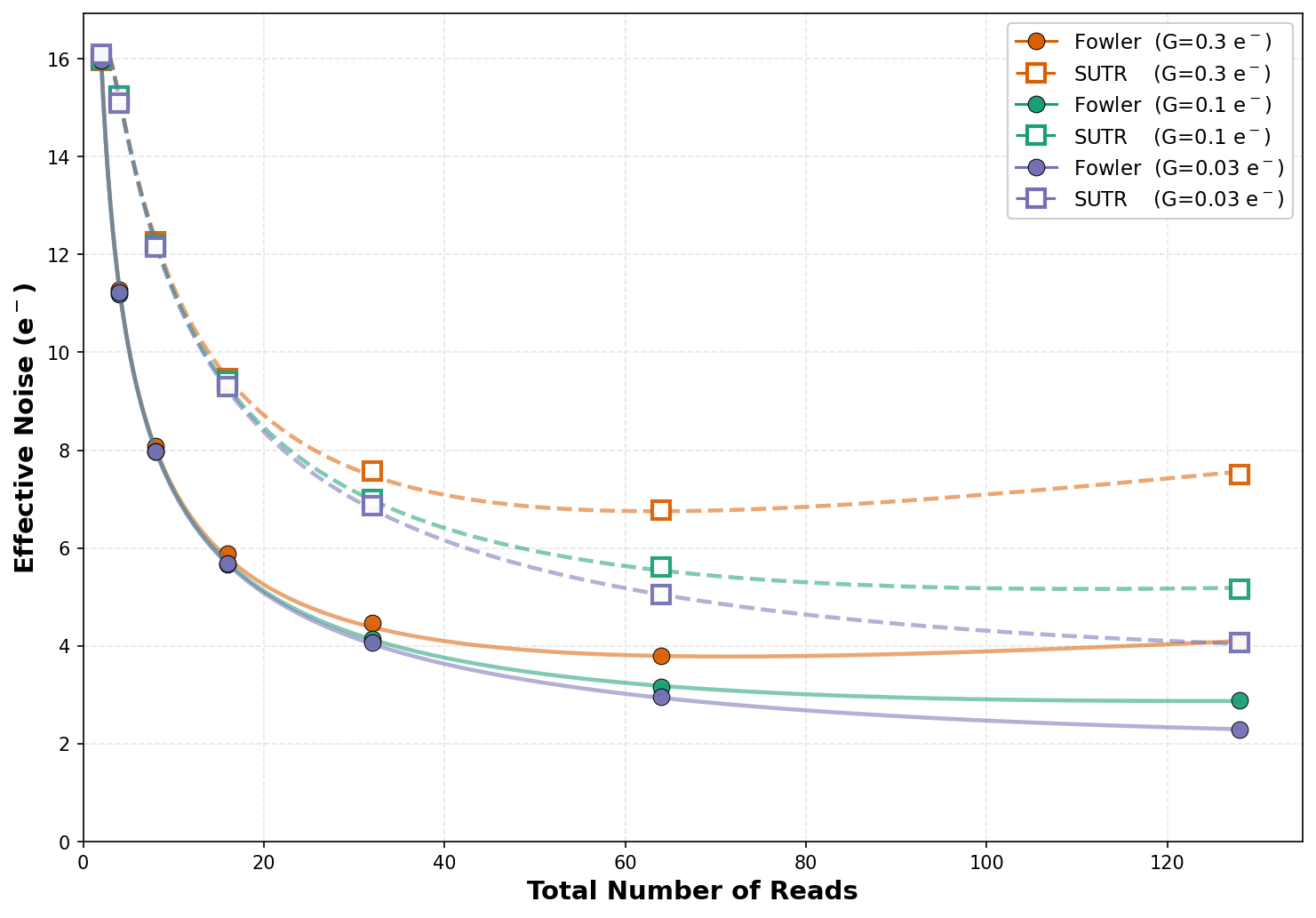}
    \caption{Simulation comparing equations \ref{eq:fowler_noise_and_mean} and \ref{eq:sutr_noise_and_mean} (lines) with numerical simulations (symbols) generating read noise of 11.1 e- RMS (16 in CDS) and averaging down the noise with different numbers of reads.}
    \label{fig:mylabel}
\end{figure}

\section{Fowler sampling mean and variance derivation}
\label{app:fowler_group1}

%Here we derive the means and variances for the case of Fowler sampling average under the non-destructive (cumulative) glow model.

\subsubsection*{Means}
For Fowler group 1, the group average is

\begin{align}
\bar{s}^{(1)}
&= \frac{1}{N}\sum_{i=1}^{N}
\left( ktc_1 + \sum_{k=1}^{i} g_k^{(1)} + r_i^{(1)} \right) \nonumber\\
&= ktc_1
+ \frac{1}{N}\sum_{i=1}^{N}\sum_{k=1}^{i} g_k^{(1)}+ \frac{1}{N}\sum_{i=1}^{N} r_i^{(1)}
\label{eq:group_1_mean}
\end{align}

The only interesting term is the glow term, where each read includes the accumulation of glow before it.  Taking the expectation

\begin{align}
\mathbb{E}\!\left[\bar{s}^{(1)}\right]
&= ktc_1
+ \frac{1}{N}\sum_{i=1}^{N}\sum_{k=1}^{i}\mathbb{E}\!\left[g_k^{(1)}\right] + \frac{1}{N}\sum_{i=1}^{N} \mathbb{E}\!\left[r_i^{(1)}\right]\\
&= ktc_1
+ \frac{1}{N}\sum_{i=1}^{N}\sum_{k=1}^{i} G \nonumber\\
&= ktc_1
+ \frac{1}{N}\sum_{i=1}^{N} iG
\label{eq:app_mean_sbar1}
\end{align}
So then 
\begin{align}
\mathbb{E}\!\left[\bar{s}^{(1)}\right] = ktc_1 + \frac{N+1}{2}\,G
\end{align}

For the second group, the group average of equation \ref{eq:fowler_read_2} is 
\begin{align}
\bar{s}^{(2)}
&= \frac{1}{N}\sum_{j=1}^{N}
\left( ktc_1 + f\,t_{\mathrm{exp}}
+ \sum_{k=1}^{N} g_k^{(1)}
+ \sum_{k=1}^{j} g_k^{(2)}
+ r_j^{(2)} \right) \nonumber\\
&= ktc_1 + f\,t_{\mathrm{exp}}
+ \sum_{k=1}^{N} g_k^{(1)}
+ \frac{1}{N}\sum_{j=1}^{N}\sum_{k=1}^{j} g_k^{(2)}
+ \frac{1}{N}\sum_{j=1}^{N} r_j^{(2)} .
\label{eq:group_2_mean}
\end{align}

The first term is the same KTC offset as before, and the second is the accumulated photoelectrons from the signal.  The third term is the fixed accumulated glow from the first exposure, $NG$ (not $(N+1)G/2$)!  Taking the expectation and using identical arguments from before, it follows that

\begin{equation}
\mathbb{E}\!\left[\bar{s}^{(2)}\right]
= ktc_1 + f\,t_{\mathrm{exp}} + \frac{3N+1}{2}\,G .
\end{equation}

\subsubsection*{Variance}

We can now examine the variance of the fowler group 1 average, Eq.~\eqref{eq:group_1_mean}, .   
\begin{align}
\mathrm{Var}\!\left(\bar{s}^{(1)}\right)
&=
\mathrm{Var}\!\left(
ktc_1
\right) + 
\mathrm{Var}\!\left(
\frac{1}{N}\sum_{i=1}^{N}\sum_{k=1}^{i} g_k^{(1)}
\right)
+
\mathrm{Var}\!\left(
\frac{1}{N}\sum_{i=1}^{N} r_i^{(1)}
\right) \\
&= \mathrm{Var}\!\left(
\frac{1}{N}\sum_{i=1}^{N}\sum_{k=1}^{i} g_k^{(1)}
\right) + \frac{\sigma_{RN}^2}{N}.
\label{eq:app_var_rn}
\end{align}

\noindent where the variance of the KTC term is zero as it's a fixed constant, and the last term follows from read noise being independent Gaussians.  To compute the glow contribution, we rewrite the double sum by swapping the
order of summation.
\begin{align}
\mathrm{Var}\!\left(
\frac{1}{N}\sum_{i=1}^{N}\sum_{k=1}^{i} g_k^{(1)}
\right)
&=
\mathrm{Var}\!\left(
\frac{1}{N}\sum_{k=1}^{N} (N-k+1)\,g_k^{(1)}
\right) \nonumber\\
&=
\frac{1}{N^2}\sum_{k=1}^{N} (N-k+1)^2\,\mathrm{Var}\!\left(g_k^{(1)}\right) =\frac{G}{N^2}\sum_{k=1}^{N} (N-k+1)^2.
\label{eq:app_var_glow_step}
\end{align}
where we used independence of the $g_k^{(1)}$.  Noting that the sequence $(N-k+1)$ for $k=1,\dots,N$ is $\{N,N-1,\dots,1\}$, the sum can be replaced by the identity $\sum_1^N  n^2 = N(N+1)(2N+1)/6$ so then the total variance is:

\begin{equation}
\mathrm{Var}\!\left(\bar{s}^{(1)}\right)
=
G\,\frac{(N+1)(2N+1)}{6N}
+
\frac{\sigma_{RN}^2}{N}.
\label{eq:app_var_sbar1_final}
\end{equation}

We can similarly assess the fowler group 2 average:
\begin{align}
\mathrm{Var}\!\left(\bar{s}^{(2)}\right)
&=
\mathrm{Var}\!\left(ktc_1\right)
+
\mathrm{Var}\!\left(f t_{exp}\right)
+
\mathrm{Var}\!\left(\sum_{k=1}^{N} g_k^{(1)}\right)
+
\mathrm{Var}\!\left(\frac{1}{N}\sum_{j=1}^{N}\sum_{k=1}^{j} g_k^{(2)}\right)
+
\mathrm{Var}\!\left(\frac{1}{N}\sum_{j=1}^{N} r_j^{(2)}\right).
\end{align}

The KTC term is again a fixed constant, and the third term is just the accumulated glow variance from the first Fowler group.  (While its value is fixed, different realizations will have different amounts of glow, so it can't be taken as a fixed constant.  This will become clear when the covariance between group 1 and 2 is computed next.)  The next two terms are computed identically to the first set, so the final expression is:

\begin{equation}
\mathrm{Var}\!\left(\bar{s}^{(2)}\right)
= f t_{exp} +
NG
+
G\,\frac{(N+1)(2N+1)}{6N}
+
\frac{\sigma_{RN}^2}{N}.
\end{equation}

\subsubsection*{Covariance}

Because the two group averages share the same glow from the first Fowler group, they are not independent and there will be a covariance between them.  Examining $\bar{s}^{(1)}$ and $\bar{s}^{(2)}$, (Eqs \ref{eq:group_1_mean} and \ref{eq:group_2_mean}), the only non-independent variables are the glow terms from the first fowler group, as expected.  This gives:

\begin{align}
\mathrm{Cov}\!\left(\bar{s}^{(1)},\bar{s}^{(2)}\right)
&=
\mathrm{Cov}\!\left(
\frac{1}{N}\sum_{k=1}^{N}(N-k+1)\,g_k^{(1)},
\ \sum_{\ell=1}^{N} g_\ell^{(1)}
\right)\\
&=
\frac{1}{N}
\sum_{k=1}^{N}(N-k+1)\,
\mathrm{Cov}\!\left(
g_k^{(1)},
\sum_{\ell=1}^{N} g_\ell^{(1)}
\right)\\
&= 
\frac{1}{N}
\sum_{k=1}^{N}(N-k+1)\,
\mathrm{Var}\!\left(g_k^{(1)}\right)\\
&=
\frac{G}{N}
\sum_{k=1}^{N}(N-k+1)\\
\end{align}

The second to last step follows since all the $g_i, g_l$'s are independent except when $k = l$, when they are identical, so the covariance can be replaced with a variance.  From similar counting arguments as before, the sum term just evaluates to $N(N+1)/2$ so the covariance is then:
\begin{equation}
\mathrm{Cov}\!\left(\bar{s}^{(1)},\bar{s}^{(2)}\right)
=
\frac{N+1}{2}\,G.
\end{equation} 

% References

\bibliography{report} % bibliography data in report.bib
\bibliographystyle{spiebib} % makes bibtex use spiebib.bst

\end{document}